\expandafter\ifx\csname natexlab\endcsname\relax\fi

\documentclass[12pt,preprint]{aastex}
\def\apjs{{\rm ApJS}}                  
\def\apjl{{\rm ApJ Let}}               

\def\Msol{\thinspace\hbox{$\hbox{M}_{\odot}$}}

\def\a4{\hsize 17.0cm \vsize 25.cm}
\newcommand{\der}[2]  { \frac{{\rm d}#1}{{\rm d}#2} }

\newcommand{\dif}     {{\rm d}}

\shorttitle{Winds from Star Clusters with a Schuster Profile}
\shortauthors{JP  et al.}

\begin{document}

\title{Young Stellar Clusters with a Schuster Mass Distribution - I: Stationary Winds}

\author{Jan Palou\v s \altaffilmark{1},
Richard W\" unsch \altaffilmark{1},
Sergio Mart\'{\i}nez-Gonz\'alez \altaffilmark{2},
Filiberto Hueyotl-Zahuantitla \altaffilmark{1},
Sergiy Silich \altaffilmark{2},
Guillermo  Tenorio-Tagle\altaffilmark{2}
}
\altaffiltext{1}{Astronomical Institute, Academy of Sciences of the Czech Republic, Bo\v cn\' \i \ II 1401-2a, Prague, Czech Republic; palous@ig.cas.cz}
\altaffiltext{2}{Instituto Nacional de Astrof\'\i sica \'Optica y
Electr\'onica, AP 51, 72000 Puebla, M\'exico}

\begin{abstract}
Hydrodynamic models for spherically-symmetric winds driven by young stellar clusters with  a generalized Schuster stellar density profile are explored. For this we use both semi-analytic models and 1D numerical simulations. We determine the properties of quasi-adiabatic and radiative stationary winds and define the radius at which the flow turns from subsonic into supersonic for all stellar density distributions. 
Strongly radiative winds diminish significantly their terminal speed and thus their mechanical luminosity is strongly reduced. 
This also reduces their potential negative feedback into their host galaxy ISM.
The critical luminosity above which radiative cooling becomes dominant within the clusters, leading to thermal instabilities which make the winds non-stationary, is determined, and its 
dependence on the star cluster density profile, core radius and half mass radius is discussed. 
  
\end{abstract}


\section{Introduction}
\label{S1}

The feedback from massive young stellar clusters to the
interstellar gas determines the natural link between the stellar and gaseous
components in galaxies. High velocity gaseous outflows driven by young stellar
clusters (star cluster winds) shape the interstellar medium (ISM) into a network
of expanding shells that engulf a hot X-ray emitting plasma \citep{2010ApJ...716..474W}. Such shells
accumulate and compress the interstellar matter often creating secondary star forming clumps within the expanding shells
\citep{2005AJ....129..393O} and massive young stellar systems in  sites of shell
collisions, as it seems to be the case of  30~Dor and other regions  in the Large Magellanic Cloud 
\citep{2013ApJ...763...56D,2009AJ....137.3599B}. 
A number of massive young stellar clusters are found in
interacting and starburst galaxies \citep{2010ARA&A..48..431P}, where the thermalization of the
kinetic energy supplied by massive stars may result in powerful,
galactic scale outflows, which link the central starburst zone of  galaxies
with the low density galactic gaseous halo and the  intergalactic medium 
\citep{1998MNRAS.293..299T,2003ApJ...597..279T}. 

Theoretical models dealing with such outflows,  assumed  either that the
energy is released in the system center, or that stars are homogeneously
distributed within the star cluster volume, as suggested in the pioneer
work  by \citet{1985Natur.317...44C}.  
The discussion of winds driven by clusters with different stellar
density distributions 
is rather incomplete: \citet{Rodriguez2007}
found an analytic non-radiative solution in the case of a power law
stellar density distributions and  compared it with 3D simulations. 
\citet{2006MNRAS.372..497J}
presented  results  from 1D numerical simulations of non-radiative
winds driven by stellar clusters with an exponential stellar density distributions.
The impact of radiative cooling on winds driven by stellar clusters with an
exponential stellar density distribution  is discussed in 
\citet{2011ApJ...743..120S},
who developed a semi-analytic method, which allows to 
localize 
the position of the singular point where the flow turns from subsonic to supersonic and calculate the run 
of all hydrodynamical variables in this case.

The observed star cluster brightness profiles, however, are different from those discussed in
all above studies. In most cases a generalized Schuster density 
profile with  $\rho_* \propto
[1+(r/R_{c})^2]^{-\beta}$ with $\beta = 1.5$, where $R_{c}$ is the core radius of the cluster stellar distribution,
provides the best fit to the empirical mass distribution in young stellar
clusters \citep{1979AZh....56..976V}. 
\citet{2001ApJ...547..317W}  used this profile to describe the distribution of pre-stellar cores (PSC) in their model of forming clusters. 
\citet{2007MNRAS.381L..40D} adopted this distribution to PSC and young stars in the Arches cluster near the center of the Milky Way.
\citet{1987ApJ...323...54E}
revealed that the generalized Schuster model with $\beta = 1.75$, provides a very good fit to the stellar densities  of young
stellar clusters in the Large Magellanic Cloud.
 Furthermore,
\citet{2002A&A...383..137M} used HST observations of
young stellar clusters in the Antennae galaxies
\citep{1999AJ....118.1551W}, and found that a King model
\citep{1962AJ.....67..471K,1966AJ.....71...64K} provides the best agreement with the observed stellar
surface densities, corresponding to a generalized Schuster model. 

The adiabatic model of  winds driven by clusters with a homogeneous stellar density distribution by  \citet[][hereafter CC85]{1985Natur.317...44C}
complemented with the effects of radiative cooling were explored by 
\citet{2004ApJ...610..226S}, 
\citet{2007ApJ...658.1196T}, \citet{2007A&A...471..579W},
\citet{2008ApJ...683..683W}, \citet{2010ApJ...708.1621T}, and
\citet{2011ApJ...740...75W}. They concluded that the importance of cooling increases for larger mass clusters. For a given cluster radius, when the cluster mass surpasses a critical value, the stationary wind solution vanishes. 

Here we extend our previous results to clusters with a generalized Schuster profile and 
discuss how it affects the
hydrodynamics of  star cluster winds. One can use the results
presented here, as a reference model, and compare them with the
observed systems. This might improve the link between observations and
model predictions, help in the interpretation of observational data and
improve our understanding of the stellar feedback and the fraction of stellar mass returned  
to the galactic ISM. 

The paper is 
organized as follows: the input star cluster model
is formulated in section 2. In section 3 we introduce the set of main equations
and present them in the form suitable for numerical integration in the
semi-analytic approach. The numerical model is presented in section 4.
Reference models are described in section 5, where the results obtained with
different methods are compared. 
We discuss separately the non-radiative solutions and  
the stationary radiative solutions through a detailed comparative description of models obtained with different energies. 
We summarize our major results in section 6. 
The non-stationary solutions including thermal instabilities will be discussed in a forthcoming communication 
\citep{Wunsch2013}.

\section{Input model}
 \label{S2}

We consider young and compact spherical clusters with constant total mass and energy 
deposition rates, $\dot{M}_{SC}$ and $L_{SC}$, and a generalized Schuster stellar  mass density distribution  \citep{1998SerAJ.158...15N}:
\begin{equation}
      \label{eq1}
\rho_{\star}(r) = \displaystyle \frac{\rho_{\star0}}{\left[1+\left(r/R_{c}\right)^2\right]^{\beta}}, 
\end{equation}
where $\rho_{\star 0}$ is the central stellar density, $R_c$ is the 
radius of the star cluster ``core'' and $\beta \geq 0$ defines the steepness of the stellar distribution. 
The cumulative mass within a given radius $r$ is then:
\begin{equation}
      \label{eq2}
M_{SC}(r) = \displaystyle \int_{0}^{r} \frac{4\pi\rho_{*0} x^2
\dif x}
{\left[1+\left(x/R_{c}\right)^2\right]^{\beta}}  =  
\frac{4\pi}{3}\rho_{*0}r^3 {}_{2}F_{1}(3/2, \beta, 5/2,
-r^2/R_{c}^2) , 
\end{equation}
where ${}_{2}F_{1}$ is the Gauss hypergeometric function,
 ${}_{2}F_{1}(3/2, \beta, 5/2,-r^2/R_{c}^2)$, hereafter  abbreviated as $F_{\beta}(r)$.

If $\beta \leq 3/2$ and $r\rightarrow \infty$, the mass of the cluster is 
infinite. 
However, if $\beta > 3/2$, the cumulative mass is finite even 
if $r \to \infty$.
In order to keep the cluster total mass finite even for $\beta \leq 3/2$, 
the stellar density distribution
(equation \ref{eq1}) must be truncated at some radius $R_{SC}$. The consideration of  the cluster radius $R_{SC}$ is justified as a consequence of environmental effects, tides etc., which remove mass from the cluster periphery.
When $\beta = 3/2$ and $R_{SC}/R_c \to \infty$, equation (\ref{eq1}) leads to 
the \citet{1962AJ.....67..471K}  
surface density distribution \citep{1998SerAJ.158...15N}.
Note,
that in the case of a homogeneous stellar mass distribution ($\beta = 0$)
the core radius $R_c$ vanishes from all formulae.

Here it is assumed, as in CC85, that the mechanical energy deposited by massive stars
and supernova explosions is thermalized in random collisions of nearby stellar
winds and supernova ejecta 
and that sources of mass ($q_m$) and energy ($q_e$) are distributed in direct
proportion to the local star density: 
\begin{eqnarray}
      \label{eq.2a}
      & & \hspace{-1.1cm} 
q_{e}(r) = \displaystyle q_{e0}\left[1+\left(r/R_{c}\right)^2\right]^{-\beta} ,
      \\[0.2cm], \label{eq.2b}
      & & \hspace{-1.1cm}
q_{m}(r) = \displaystyle q_{m0}\left[1+\left(r/R_{c}\right)^2\right]^{-\beta} ,
\end{eqnarray}
where the normalization constants $q_{e0}$ and $q_{m0}$ are:
\begin{eqnarray}
      \label{eq.3a}
      & & \hspace{-1.1cm} 
q_{e0} = \displaystyle 3 L_{SC}/4\pi R_{c}^3 F_{\beta}(R_{SC}),
      \\[0.2cm]     \label{eq.3b}
      & & \hspace{-1.1cm}
q_{m0} = \displaystyle 3 \dot{M}_{SC}/4\pi R_{c}^3  F_{\beta}(R_{SC}).
\end{eqnarray}

\section{Semi-analytic approach}
\label{sec:semianl}

\subsection{\it Basic equations}
\label{S3}

The  hydrodynamic equations for the steady state, spherically symmetric flows
driven by clusters with energy and mass deposition rates $q_e(r)$ and 
$q_m(r)$ are (see, for example, \citealp{1971ApJ...165..381J, 
1985Natur.317...44C, 2000ApJ...536..896C, 2004ApJ...610..226S}):
\begin{eqnarray}
      \label{eq4a}
      & & \hspace{-1.0cm}
\frac{1}{r^2} \der{}{r}\left(\rho u r^2\right) = q_m ,
      \\[0.2cm]
      \label{eq4b}
      & & \hspace{-1.0cm}
\rho u \der{u}{r} = - \der{p}{r} - q_m u,
      \\[0.2cm]
     \label{eq4c}
      & & \hspace{-1.0cm}
\frac{1}{r^2} \der{}{r}{\left[\rho u r^2 \left(\frac{u^2}{2} +
\frac{\gamma}{\gamma - 1} \frac{p}{\rho}\right)\right]} = q_e - Q ,
\end{eqnarray}
where $u$, $p$, and $\rho$ are the gas outflow velocity, thermal pressure
and density, respectively, $\gamma$ (=5/3) is the ratio of the specific heats,
$Q = n_e n_i \Lambda(T,Z)$ is the cooling rate, $n_e$, $n_i$ are the number densities of electrons and ions, and $\Lambda(T,Z)$ is the 
cooling function, which depends on the gas temperature $T$ and metallicity 
$Z$. We use the equilibrium cooling function for optically thin plasma 
obtained by 
\citet{1995MNRAS.275..143P}. 
In all calculations the metallicity of the plasma 
was set to the solar value.  Hereafter we relate the energy and the mass 
deposition rates, $L_{SC}$ and ${\dot M}_{SC}$, via the equation:
\begin{equation}
      \label{eq5}
L_{SC} = {\dot M}_{SC} V^2_{A\infty} / 2 ,
\end{equation}
and assume that the adiabatic wind terminal speed, $V_{A\infty}$, is 
constant. For a known star cluster mechanical luminosity $L_{SC}$, the parameter 
$V_{A\infty}$ defines the mass deposition rate.

The integration of the mass conservation equation (\ref{eq4a}) yields:
\begin{equation}
      \label{eq6}
\rho u r^2 = q_{m0} r^3 F_{\beta}(r)/3 + C .
\end{equation}
If the density and the velocity of the flow in the star cluster center
are finite, the constant of integration must be zero: $C = 0$.  Using this 
expression and taking the derivative of equation (\ref{eq4c}), one can 
present the main equations in a form suitable for numerical integration:
\begin{eqnarray}
 \label{eq7a}
      & & \hspace{-1.1cm} 
\der{u}{r}  = \frac{\displaystyle (\gamma-1)(q_{e}-Q)
+(\gamma+1)q_{m}u^2/2
-2c^2\rho u/r}
{\displaystyle   \rho(c^2-u^2)}, 
      \\[0.2cm]     \label{eq7b}
      & & \hspace{-1.1cm}
\der{p}{r} = - \rho u \der{u}{r} - q_m u \, ,
      \\[0.2cm] \label{eq7c}
      & & \hspace{-1.1cm}
\rho = \frac{q_{m0}r}{3 u} F_{\beta}(r) \, ,
\end{eqnarray}
where $c$ is the local speed of sound, $c^2 = \gamma p / \rho$.  

Given a cluster radius $R_{SC}$, outside of which there are no sources of mass and energy, the set of the main equations for $r > R_{SC}$ is:
\begin{eqnarray}
 \label{eq9a}
      & & \hspace{-1.1cm} 
\der{u}{r}  = \frac{(\gamma-1) r Q + 2 \gamma p u}{r \rho(u^2-c^2)}, 
      \\[0.2cm]     \label{eq9b}
      & & \hspace{-1.1cm}
\der{p}{r} = - \frac{{\dot M}}{4 \pi r^2} \der{u}{r}  \, ,
      \\[0.2cm] \label{eq9c}
      & & \hspace{-1.1cm}
\rho = \frac{{\dot M}}{4 \pi u r^2}  \, ,
\end{eqnarray}
where $\dot M$ is the flux of mass through the star cluster surface.

\subsection{\it Integration procedure}
\label{sec:windsol}

As a consequence of equation (\ref{eq7c}) the central gas density $\rho_0 $ is non zero and remains finite 
only if the flow velocity at $r = 0$ is 
zero, and grows linearly with radius near the center. The derivatives of 
the wind velocity and pressure in the star cluster center then are: 
\begin{eqnarray}
 \label{eq8a}
      & & \hspace{-1.1cm} 
\der{u}{r} = \left[(\gamma-1)(q_{e0} - Q) - 2 q_{m0} c_0^2 / 3 \right]  / 
      \rho_0  c^2_0 \,  ,
      \\[0.2cm]     \label{eq8b}
      & & \hspace{-1.1cm}
\der{p}{r} = 0 \, ,
\end{eqnarray}
where $c_0$ is the sound speed in the star cluster center. 
It is interesting to note, that these 
relations are identical to those, obtained 
for the top-hat or homogeneous and for  the exponential stellar density distribution by \citet{2004ApJ...610..226S,2011ApJ...743..120S}, and that they do not depend on 
the selected value of $\beta $. We make use of these equations in 
order to move from the center and start the numerical integration. 

In the radiative wind model, the central gas density $\rho_0$ and the central temperature $T_0$ are 
related through the equation
\citep{Sarazin1987,2004ApJ...610..226S}:
\begin{equation}
      \label{eq9}
n_{0} = q_{m0}^{1/2}\left[\frac{V_{A,\infty}^{2}/2 - 
            c_{0}^{2}/(\gamma -1)}{\Lambda (Z,T_{0})}\right]^{1/2} \, 
\end{equation}
where
$n_0 =\rho_0 / \mu m_p$ is the central number density of ions   
and $\mu m_p$ is the average mass per ion. 
Thus the central temperature $T_0$ is the only parameter which selects the 
solution from a branch of possible integral curves.

Searching for the physical wind solution, two cases may occur. In the first one, it is possible to find the unique solution which passes through the singular point $R_{sg}$, where both, the numerator and the denominator in equation 
(\ref{eq7a}) vanish  inside of the cluster: $R_{sg} < R_{SC}$. At this point, the subsonic flow  in the region $r < R_{sg}$ changes to a supersonic  flow in the region $r > R_{sg}$,  thus the singular point is also the sonic point. 
The presence of $R_{sg}$ inside the cluster allows one to select the 
value of the central temperature and the unique wind solution. The position of the singular point $R_{sg}$ can be calculated using the method described by 
\citet{2011ApJ...743..120S}, where
the formulae
which define the hydrodynamic variables and the derivative of the flow
velocity at $R_{sg}$, and which allow to pass it 
in the semi-analytic calculations, are given.

In the other case, if the singular point does not exist inside the cluster, the transition from a subsonic to a supersonic flow occurs abruptly at $R_{SC}$, where the stellar density  changes discontinuously and where the velocity gradient is infinite,  CC85 and \citet{2000ApJ...536..896C}. In this case the numerator and denominator of equation (12) are both positive when one approaches $R_{SC}$ from the inside, and both negative when one approaches it from the outside.

\section{Numerical simulations}

We perform 1D numerical simulations to complement the semi-analytical calculations and confirm the results.
Full numerical simulations are also required in order to find the hydrodynamic solution in the case of very energetic clusters, where strong radiative cooling promotes thermal instabilities and inhibits a stationary solution. 
The models are calculated with the finite-difference Eulerian hydrodynamic code FLASH \citep{2000ApJS..131..273F}. Here we perform the 1D numerical simulations assuming spherically symmetric clusters.
 The calculation of radiative cooling within the computational domain and its impact on the time-step is computed following 
\citet[][hereafter GTT07]{2007ApJ...658.1196T} 
and 
\citet[][hereafter W08]{2008ApJ...683..683W}, 
where the cooling routine uses the \citet{Raymond1976}  cooling function updated by \citet{1995MNRAS.275..143P}. 

In GTT07 and W08 the flow was modeled considering a continuous replenishment of internal energy and mass in all cells within the cluster volume at rates $q_e (r)$ and $q_m (r)$, respectively. The dependence of these quantities on radius is given by equations (\ref{eq.2a}) and (\ref{eq.2b}) with the normalization constants given by equations (\ref{eq.3a}) and (\ref{eq.3b}).  
$\dot{M}_{SC}$ is related to $L_{SC}$ as given by equation (\ref{eq5}) through the constant 
$V_{A\infty}$, thus defining the mass flux in the cluster wind. Small values of  $V_{A\infty}$ 
mean that the mass in winds of individual stars is loaded by additional mass from 
the parental cloud.  
At every time-step energy and mass are inserted within the cluster volume following the procedure described in GTT07 and W08.

\section{Results and discussion}

\subsection{Non-radiative winds}
 
For a stationary wind the
total energy flux $L(r)$ through a sphere of radius $r$ is:
\begin{equation}
\label{Lr}
L(r) \equiv 4\pi r^2 \rho u (\frac{u^2}{2}+\frac{\gamma}{\gamma-1}\frac{p}{\rho}) \ .
\end{equation}
 In the non-radiative case, this has to be equal to the energy 
\begin{equation}
\label{LSCr}
L(r) = \int_0^r 4\pi x^2 q_e dx = 4\pi r^3 
F_{\beta}(r)/3
\end{equation}
inserted by stars into a sphere with the same radius. 
At the  sonic point $R_{son}$, the wind velocity
fulfills $u(R_{son}) = c(R_{son})$ which together with equations
(\ref{Lr}) and (\ref{LSCr}) and the sound speed definition $c^2 = \gamma p/\rho$
yields
\begin{equation}
\label{LrLSCr}
3\rho c^3 (\gamma+1) = 2 (\gamma -1) q_{e0} R_{son}
F_{\beta}(R_{son}).
\end{equation}
Inserting
the continuity equation (\ref{eq7c}) into (\ref{LrLSCr}) and utilizing 
$V_{A\infty}^2
= 2q_{e0}/q_{m0}$ one gets
\begin{equation}
c^2(R_{son}) = \frac{\gamma-1}{\gamma+1} V_{A\infty}^2.
\label{c-rson}
\end{equation}
This implies that in the adiabatic case ($Q = 0$), the sound speed at the
sonic point $R_{son}$,  
is exactly one half of the terminal wind velocity {\bf $V_{A\infty}$}, if $\gamma = 5/3$ \citep[see][]{2000ApJ...536..896C}.

\begin{figure}
\plotone{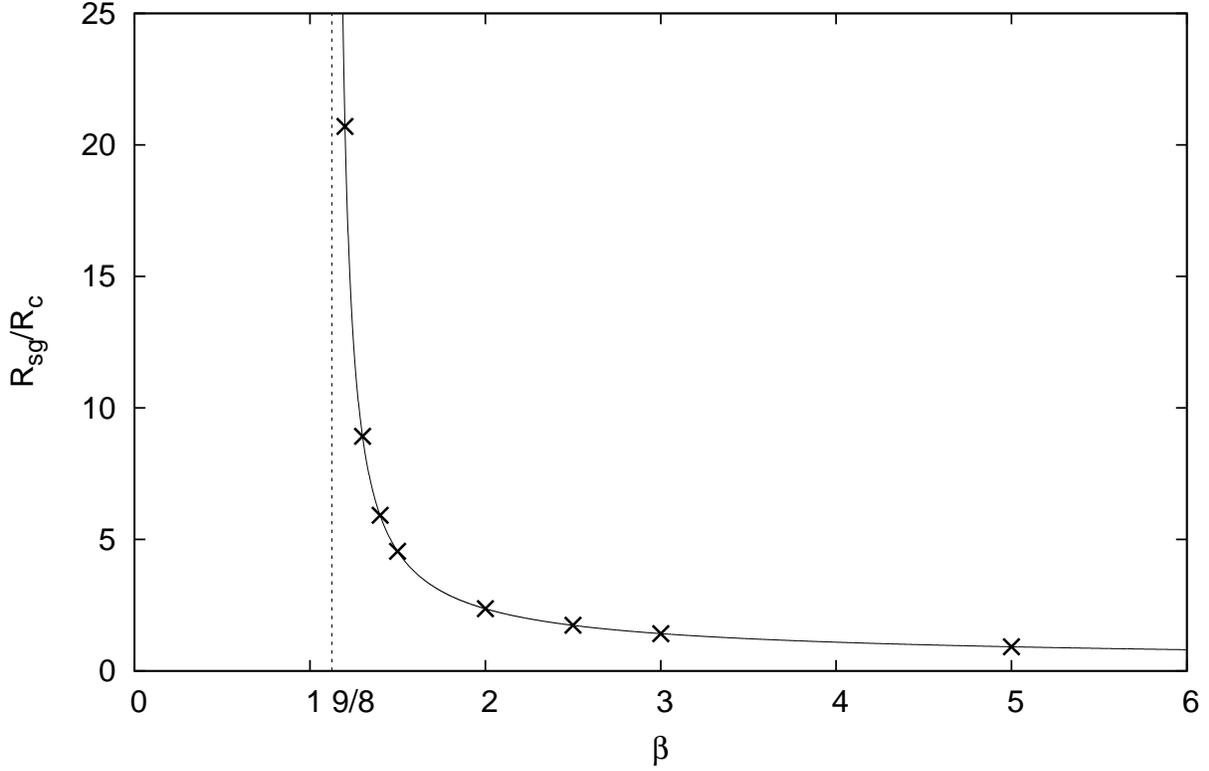}
\caption{The position of the singular point $R_{sg}$ normalized by the core radius $R_c$ as a function of $\beta$. Solid line shows the solution of the equation (\ref{eq:rsonbeta}), x symbols give the position of the sonic points measured in 1D hydrodynamical simulations. The vertical dotted line marks the limiting $\beta$ value $\beta_{crit}$ for which the singular point goes to infinity. 
}
\label{fig:rsonbeta}
\end{figure}

When the transition to the supersonic regime occurs inside the cluster,  the  denominator and the numerator of  equation (\ref{eq7a}) are both equal to zero and $R_{son} = R_{sg}$. 
In the non-radiative case ($Q = 0$) it leads to:
\begin{equation}
(\gamma-1)q_e + (\gamma+1)q_m \frac{c}{2} - \frac{2}{3}q_{m0} c^2
F_{\beta}(R_{sg})
= 0.
\end{equation}
Equations (\ref{eq.2a}), (\ref{eq.2b}) and (\ref{c-rson}) then lead to   
the algebraic equation for $R_{sg}$
\begin{equation}
\label{eq:rsonbeta}
\left[1+\left(\frac{R_{sg}}{R_c}\right)^2\right]^{-\beta}
=
\frac{4}{3(5\gamma-3)}
F_{\beta}(R_{sg})
\ .
\end{equation}
The solution of equation (\ref{eq:rsonbeta}) can be found numerically. 
It is a function of only one parameter ($\beta$).
The position of the adiabatic singular point for all clusters 
with a Schuster stellar density profile is shown in Figure \ref{fig:rsonbeta}, where 
the solution of equation
(\ref{eq:rsonbeta}) is also compared to the sonic point positions measured in 1D hydrodynamic
simulations. There is an excellent agreement between the two methods. 
The dependence on the sources (stellar) density
distribution alone was already claimed by \citet{2006MNRAS.372..497J}.  

It is important to
note, that in the adiabatic case (when Q = 0) there is a critical value of $\beta$ ($\beta_{crit} = 1.125$) such that for $\beta \le \beta_{crit}$ the singular point $R_{sg} \to \infty$. 
This implies, that in clusters with a shallow stellar density
distributions ($\beta \le \beta_{crit}$), 
the transition to the supersonic regime occurs at infinity 
(see Figure \ref{fig:rsonbeta}),
or if truncated, at the star cluster surface $R_{SC}$. On the other hand,  in clusters with steeper
density gradients the transition to the supersonic regime occurs inside the
cluster, if the star cluster is sufficiently large:
$R_{SC} > R_{sg}$. 

\subsection{\it Reference models}

\begin{figure}[htbp]
\includegraphics[angle=-90,width=1.20\textwidth]{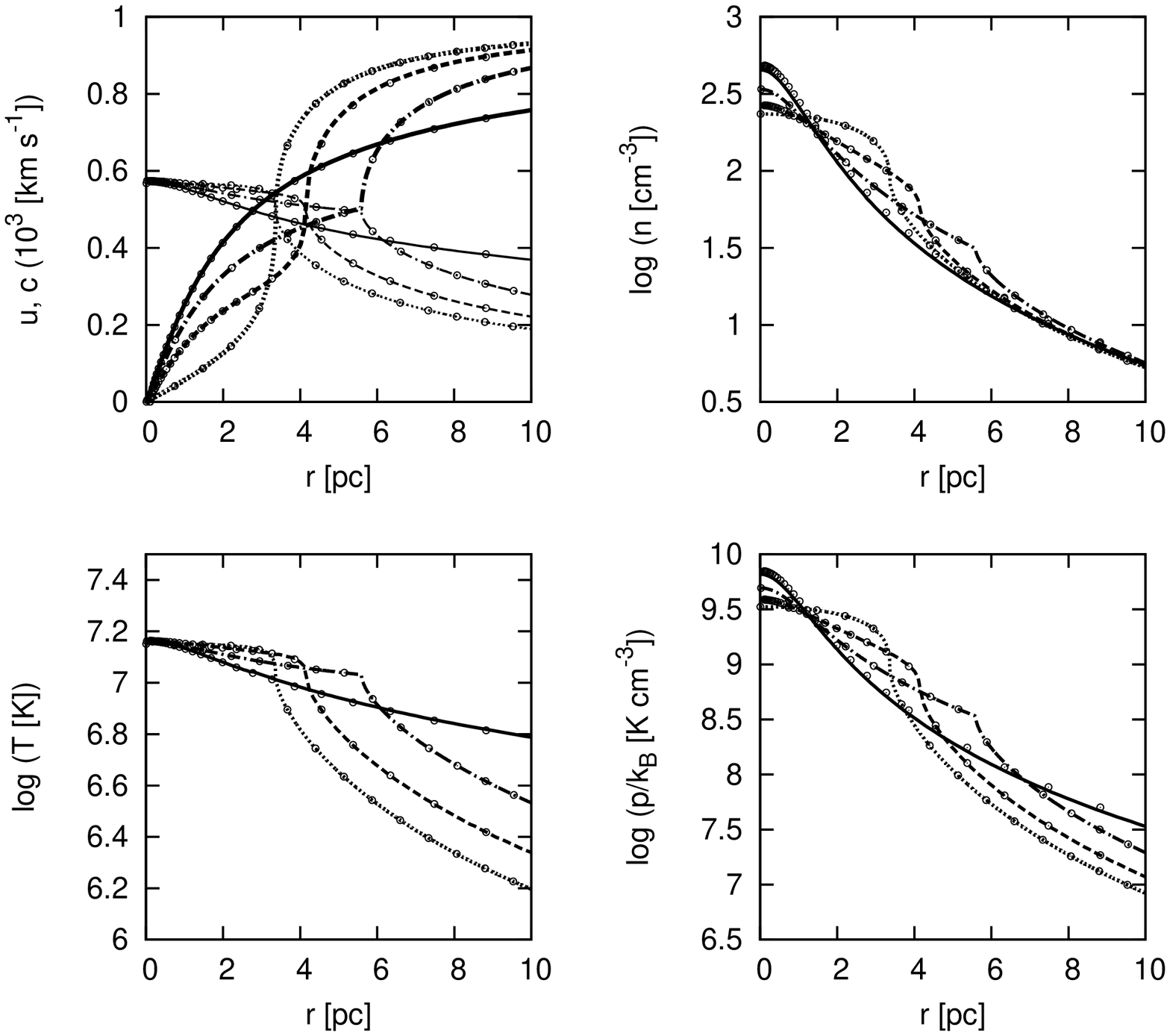}
\caption{Stationary wind solution for quasi-adiabatic winds. The wind
velocity $u$ - thick lines and the sound speed $c$ - thin lines (upper left), the density $n$ (upper right), the temperature $T$ (lower left) and the thermal pressure $p/k_B$ (lower right) are shown. $k_B$ is the Boltzmann constant. The lines present the results from the semi-analytic calculations and the circle symbols give the results of 1D hydrodynamical simulations: Model I (dotted lines), Model II (dashed lines), Model III (dash-dotted lines), Model IV (solid lines). The transition to a supersonic regime occurs at radii, 
where on the left upper panel the thin and
thick lines of the corresponding model cross.}  
\label{fig2}
\end{figure}

\begin{table}[htp]
\caption{\label{tab1} Reference models}
\begin{tabular}{c c c c c c c c}
\hline\hline
Model  & $\beta$ & Core radius & Cluster radius & Sonic radius \\
&       & $R_c$(pc)  & $R_{SC}$(pc)   & $R_{son}$(pc)  \\
(1) & (2) & (3) & (5) \\\hline
I  &  0 & ...  & 3.36  & 3.36  \\
II  &  1.0 & 1.176  & 4.14 & 4.14 \\
III  &  1.5 & 1.176  & 5.59 & 5.34  \\
IV  &  2.0 & 1.176  & $\infty$ & 2.78  \\
\hline\hline
\end{tabular}
\end{table}

The input parameters for our reference models are presented
in Table 1. All reference clusters (Models I, II, III and IV) include radiative cooling (see equations 9, 12 and 15) and have the same half-mass radius  as
the exponential model of \citet{2011ApJ...743..120S}: $R_{hm} = 2.67$~pc and the same adiabatic wind terminal speed:  $V_{A \infty} = 1000$~km s$^{-1}$.
For Model I, which has the flat top-hat stellar density profile ($\beta = 0$), the value of the cut-off radius is $R_{SC} = 3.36$~pc. 
The core radius ($R_c = 1.176$~pc) was chosen so that for $\beta = 2$ the half mass radius is $R_{hm} = 2.67$~pc with $R_{SC} \rightarrow \infty$. 
We use the same value of $R_c = 1.176$~pc in models II and III, what  leads to the star cluster radii $R_{SC} = 4.14$ pc and $R_{SC} = 5.59$ pc, respectively.  

First, we computed the hydrodynamic variables in our
reference Models I - IV as functions of $r$ for  a mechanical luminosity $L_{SC} = 3 \times 10^{40}$~erg/s, 
which  is typical for young stellar clusters with
masses {\bf $\sim$} $10^6$~\Msol. For this mechanical luminosity  radiative cooling is not important and all the models behave quasi-adiabatically.
Figure \ref{fig2} shows the distributions of the flow velocity and sound speed 
(upper left panel),  density (upper right panel), temperature (lower left panel) and 
thermal pressure (lower right panel).
The lines show the results from the semi-analytic 
calculations, whereas circle symbols give the hydrodynamic variables obtained in 1D simulations.
 The results from the semi-analytic and 1D simulations are in an excellent agreement.

If the transition to the supersonic regime occurs at the star cluster surface,
as in the case of Models I and II, the velocity jumps abruptly from subsonic to supersonic, and the density, temperature and pressure decrease sharply at the cluster edge. If the sonic point resides inside the cluster, as in Models III and IV, the transitions are much more gradual and smooth. 

The slowest radial decrease in pressure inside the cluster is observed in Model I with the top-hat stellar density profile. This results in a very slowly rising radial velocity inside the cluster followed by a very sharp transition to the supersonic flow at the cluster edge. There, at $R_{SC}$, due to a large gradient in the  wind velocity, the wind density $n$, the wind temperature $T$, and the wind pressure $p$ drop sharply.

A slightly larger pressure gradient and the related velocity growth in the central zones is observed for Model II, even larger for Model III and the largest one for Model IV, due to the growing steepness of the stellar density distribution. In the case of Models I and II the density $n$ of these winds also goes down  abruptly at $R_{son} = R_{sg}$, whereas in Models III and IV the density decreases in a more continuous way.   

The abrupt jump in velocity at $R_{SC}$ in the case of Model I, leads to the fast decrease in temperature and in pressure at this point whereas the wind temperature in  Model IV decreases slowly, making this wind the hottest and largest pressure at large distances from the center.  Model IV, which has the steepest slope in the stellar distribution, also has  the largest velocity gradient and the largest density and pressure in the center. The density, temperature and pressure of Model IV at the star cluster edge are below  other models,  while further out, at $r \sim 10 \times R_c$, are above.
Thus, the distribution of all hydrodynamical variables depends significantly on the stellar density distribution.         

\subsection{\it The impact of strong radiative cooling}

\begin{figure}[htbp]
\includegraphics[angle=-90,width=1.20\textwidth]{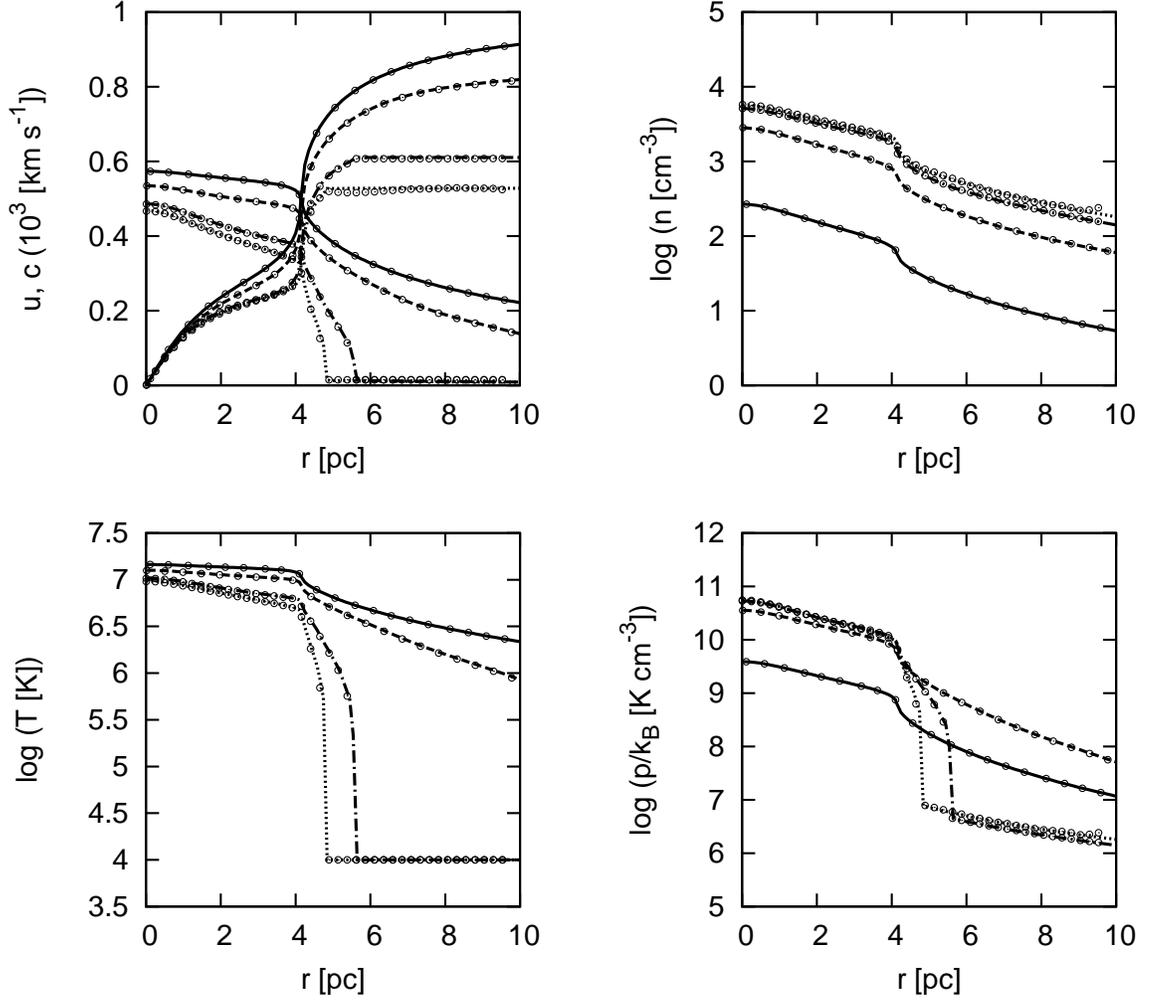}
\caption{Stationary wind solution for strongly radiative winds with stellar distribution parameters of Model II. 
Wind velocity $u$ and local sound speed $c$ (upper left), the wind density $n$ (upper right), the wind temperature $T$ (lower left) and the wind pressure $p/k_{B}$ (lower right) as a function of radius for clusters with $L_{SC} = 3.0 \times 10^{40}$ erg s$^{-1}$ (solid lines), $3.0 \times 10^{41}$~erg s$^{-1}$ (dashed lines),  $5.25 \times 10^{41}$~erg s$^{-1}$ (dashed-dotted lines) and  $5.82 \times 10^{41}$~erg s$^{-1}$ (dotted lines). Lines give results of the semi-analytical calculations and circles show 1D hydro simulations.
}
\label{fig3}
\end{figure}

\begin{figure}[htbp]
\includegraphics[angle=-90,width=1.20\textwidth]{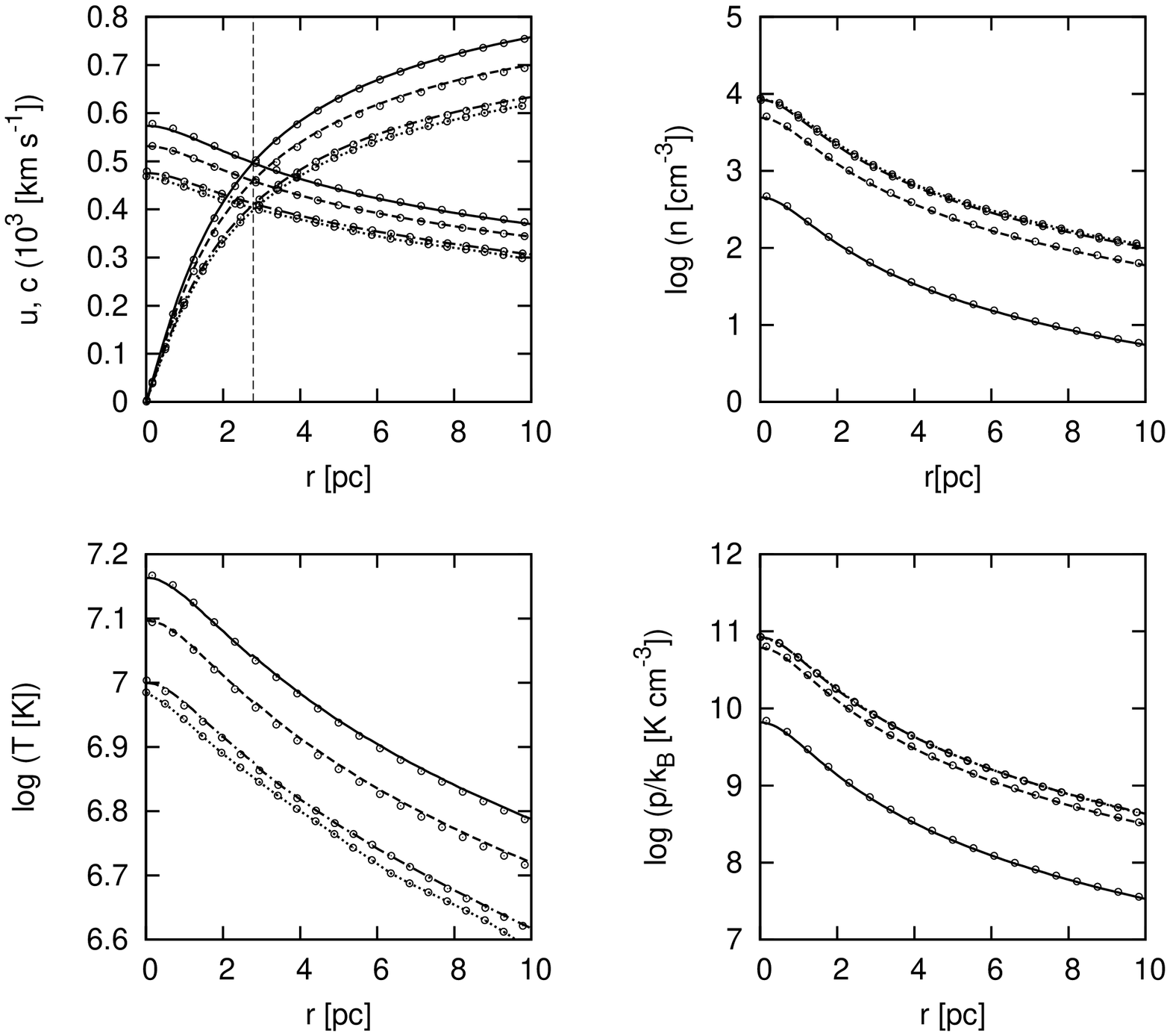}
\caption{Stationary wind solution for strongly radiative winds with stellar distribution parameters of Model IV. 
Wind velocity $u$ and local sound speed $c$ (upper left), the wind density $n$ (upper right), the wind temperature $T$ (lower left) and the pressure $p/k_B$ (lower right) as a function of radius for clusters with $L_{SC} = 3.0 \times 10^{40}$~erg s$^{-1}$ (solid lines), $3.0 \times 10^{41}$~erg s$^{-1}$ (dashed lines),  $4.6 \times 10^{41}$~erg s$^{-1}$ (dash-dotted lines) and $4.8 \times 10^{41}$~erg s$^{-1}$ (dotted lines). Lines give results of the semi-analytical calculations and circles show 1D hydro simulations. The vertical dashed line in the upper left panel shows the position of the sonic point for the adiabatic case (see Figure 1).
}
\label{fig4}
\end{figure}

The impact of radiative cooling on the flow increases with an increasing wind density.  Cooling is proportional to the square of density, which is linearly proportional to the total mass of the cluster. Thus cooling becomes more and more important as one considers a larger cluster mass. In order to explore the impact of cooling on the cluster wind behavior, the cluster mechanical luminosity was set to $L_{SC} = 3 \times 10 ^{40}, 3 \times 10 ^{41}, 5.25 \times 10^{41}$ and $5.82 \times 10^{41}$ erg s$^{-1}$ in Model II and to $L_{SC} =  3 \times 10 ^{40},  3 \times 10^{41}, 4.6 \times 10^{41}$ and $4.8 \times 10^{41}$ erg s$^{-1}$ in Model IV. These values were chosen so that the impact of cooling is clearly visible, however, the winds still remain stationary.    

Radiative wind solutions of Model II are plotted in Figure \ref{fig3}, and of Model IV in Figure \ref{fig4}. In both figures,  the results of semi-analytical calculations together with 1D hydrodynamical simulations are given. Notice again the excellent agreement between the solutions obtained with the two very different approaches.
 
For a total mechanical luminosity $L_{SC} = 3 \times 10^{40}$~erg s$^{-1}$  radiative cooling is not important leading to a  quasi-adiabatic solution  as  already presented in Figure \ref{fig2}.
However, for larger cluster masses radiative cooling starts to play a major role.  

Cooling removes a fraction of the inserted mechanical energy leading to an overall decrease of the wind temperature and thus of the local sound speed. The wind velocity also decreases in this denser plasma, despite the fact that  
the 
wind thermal pressure grows with increasing cluster luminosity. 
At some distance from the center the  wind cools down to $3 \times 10^5$ K. There, cooling speeds up and the temperature quickly reaches its lowest allowed value, which is $10^4$ K. This is similar to the behavior of cluster winds with  a top-hat and an exponential density profile described by 
\citet{2004ApJ...610..226S, 2011ApJ...743..120S},
GTT07 and W08. For larger $L_{SC}$ this happens closer to the cluster center. This dramatic temperature decrease leads to a sharp decrease in pressure. 

\begin{figure}[htbp]
\includegraphics[angle=0,width=0.7\textwidth]{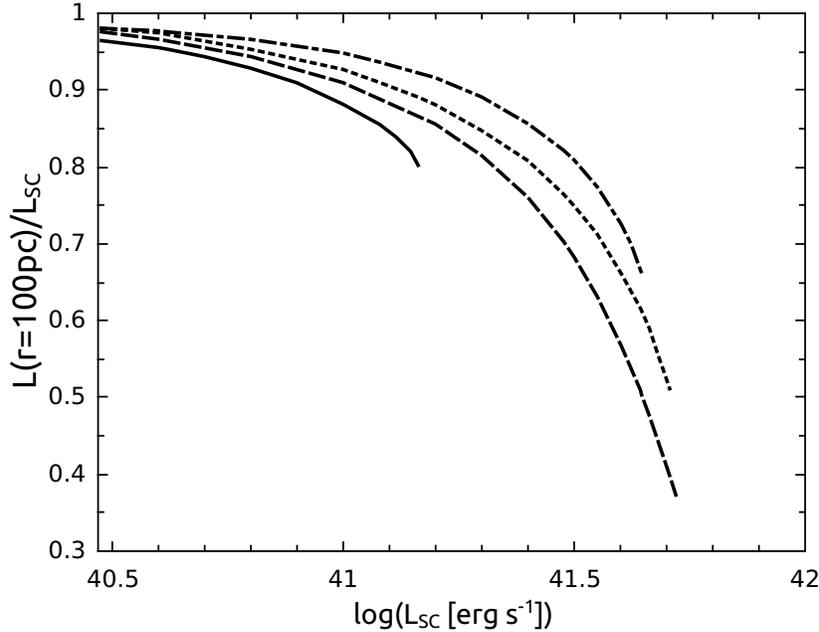}
\caption{The fraction of inserted energy retained by stationary winds as a function of the star cluster luminosity, for $\beta$ = 0, 1, 1.5 and 2. (solid, long dashed, short dashed and dash dotted lines, respectively)}
\label{fig5new}
\end{figure}

Thus, there is a situation in which  more massive clusters present 
near to their centers 
a larger pressure due to a density enhancement, while 
at larger distances  their thermal pressure drops more rapidly due to strong radiative cooling.
The  mechanical luminosity of radiative winds is strongly diminished leading to a decrease in the wind terminal speed.   
The fraction of the total energy flux retained by the wind decreases with increasing cluster mechanical luminosity as progressively a larger fraction of the deposited energy is radiated away.
Figure \ref{fig5new}  shows  the flux of total energy (see equation \ref{Lr}) through a sphere of radius 100 pc as a function of the star cluster deposited mechanical luminosity $L_{SC}$.  The right most points on the curves mark the largest mechanical luminosity for which a stationary wind solution exists. The largest loss of energy in stationary winds occurs for Models II and III with $\beta = 1$ and $\beta = 1.5$, respectively. For steeper or flatter  stellar density profiles the stationary winds vanish at lower mechanical luminosities. 

Note also that stationary winds from 
clusters with a steeper stellar density profiles retain a larger fraction of their deposited energy. This is mainly  due to the resultant slope of the wind density distributions, which closely follows that of the stars, and related wind speeds. As we can see in Figures \ref{fig3} and \ref{fig4}, inside of the star cluster at $R = 2$ pc, the density is higher in the case of model II with $\beta = 1$ compared to model IV with $\beta = 2$, however the wind speed is lower  in model II compared to model IV.  
Steeper stellar distributions lead to smaller sizes of the strongly radiative high density central regions with winds of high velocity. This leaves less time for cooling in the case of steep than is the case of shallow stellar density profiles.

\subsubsection{The Critical Line - semi-analytical solutions and 1D hydrodynamic simulations}

For larger mass clusters or larger mechanical luminosities, one would soon reach the point above which a stationary wind solution does not
exist.  Physically this means that the hot gas inside the cluster is so dense that the energy deposition by massive stars cannot
balance the loses due to expansion and radiative cooling and a large fraction of the gas inevitably cools down.    
As soon as the temperature drops to several times
$10^{5}$~K radiative cooling becomes extremely fast due to free-bound and
bound-bound transitions and some regions cool down very quickly to $\sim 10^4$~K
(we do not allow the gas to cool below this temperature assuming that there are
enough UV photons to maintain the gas warm and ionized). Consequently, these
warm regions are compressed into dense clumps by the surrounding hot gas where
the pressure is initially approximately three orders of magnitude larger. 

For a given combination of cluster parameters $\beta$, $R_c$ and $R_{SC}$, there
is a critical luminosity, $L_{crit}$, separating the region of stationary winds
from the region where thermal instabilities occur within the cluster volume   what leads to clump formation and to non-stationary outflows. The critical luminosity
can be determined either by the semi-analytical code by searching for the $L_{SC}$
above which the solution of equations (12)-(14) does not exist, or by 1D
hydrodynamic simulations. 

In the case of semi-analytical calculations, we distinguish the two following situations: a) Clusters with $\beta=0$, clusters with $\beta > \beta_{crit}$, and compact clusters with $\beta \leq \beta_{crit}$ (e.g. for $\beta=1$ and $R_{SC}/R_{c} \lessapprox 8$ and $R_{SC}/R_{c} \lessapprox 4.3$ for $\beta=0.5$ ). For these clusters the criterion discussed in Tenorio-Tagle et al. (2007) for the homogeneous case, was used. i.e.  the transition to the thermally unstable solutions occurs soon after the central temperature $T_{pm}$ drops to the value  which corresponds to the maximum of the central pressure. This temperature  is irrespective of  the values of $\beta$, $R_c$ and $R_{SC}$ and depends only on {\bf $V_{A\infty}$} and $Z$. $T_{pm}$ can be calculated by solving the equation:
\begin{equation}
    \label{eq27}
1-\frac{q_{m0} \mu_a T_{pm}}{2 \mu_i n_0^2 \Lambda (T_{pm},Z)} 
\left[\frac{c_0^2}{(\gamma-1)T_{pm}}+\frac{n_0^2}{q_{m0}}\der{\Lambda (T_{pm},Z)}{T_{pm}} \right]=0,
\end{equation}
which is equivalent to equation (7) of \citet{2009ApJ...700..931S} 
with the heating efficiency, $\eta=1$.\\
All these cases have in common the fact that {\bf $R_{sg}$} remains at its adiabatic position.
b) More extended clusters with $\beta \leq \beta_{crit}$ (e.g. for $\beta$ = 1 with $R_{SC}/R_{c} \gtrapprox 8$ and for $\beta=0.5$ and $R_{SC}/R_{c} \gtrapprox 4.3$). In these cases, strong radiative cooling forces $R_{son}$ to detach from its adiabatic position, moving towards the center as one considers more massive clusters. Thus, the run of the hydrodynamical variables changes qualitatively, promoting the onset of thermal instabilities. Therefore, $L_{crit}$ is defined as the value for which $R_{son}$ begins to detach from its adiabatic position.  
The two semi-analytical criteria have been combined to define a unique semi-analytical curve for $L_{crit}$ (see Figure 6).

In the case of 1D hydrodynamical simulations, we vary $L_{SC}$ and use the
bisection method to search for the largest $L_{SC}$ for which no zones with
temperature smaller than $10^5$~K appear inside of the cluster or in the case of Models III and  IV in the region $r < 1.1R_{sg,adia}$,
where $R_{sg,adia}$ is the singular point for clusters in the adiabatic regime
with a given $\beta$ (calculated using equation (26)).
  
\begin{figure}[htbp]
\includegraphics[angle=0,width=1.0\textwidth]{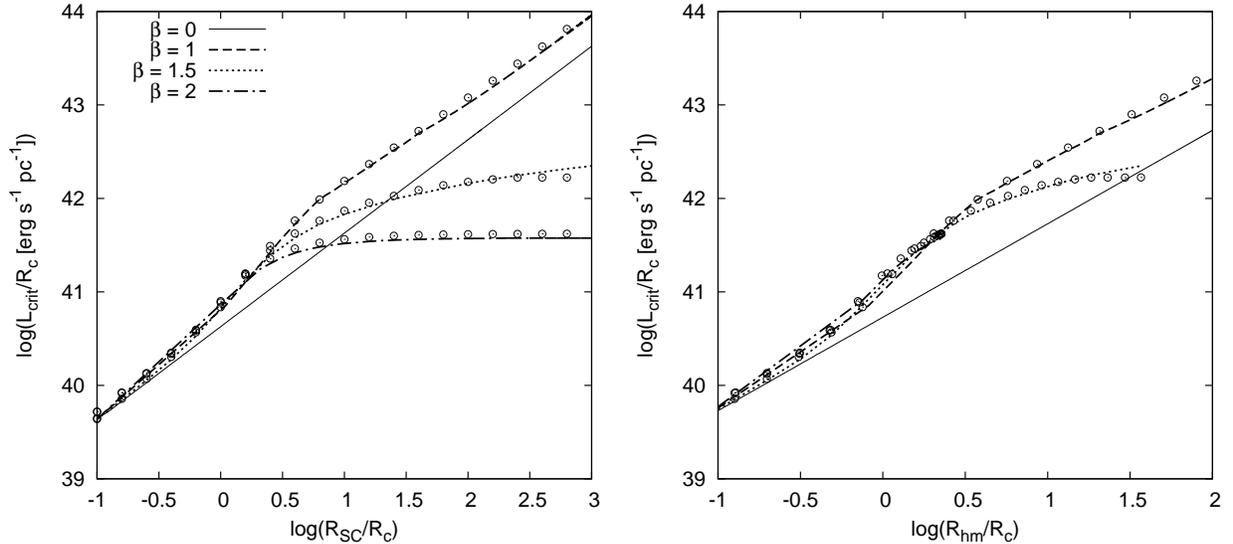}
\caption{$L_{crit}/R_c$ as a function of $R_{SC}/R_c$ (left panel) and $R_{hm}/R_{c}$ (right panel) for $\beta=0$ (solid line), $\beta=1$ (dashed line), $\beta=1.5$ (dotted line) and $\beta=2$ (dash - dotted line). 
Lines correspond to the semi-analytic results and 
open circles give the results of the 1D hydro-simulations.}  
\label{fig5}
\end{figure}

Figure \ref{fig5} compares the results of the semi-analytical calculations with numerical results for different $\beta$ values.  
Since $L_{crit}$ is directly proportional to the size of the cluster, as shown by W08 for clusters with the top-hat profiles,  the critical luminosity $L_{crit}$ is normalized to the star cluster core radius $L_{crit}/R_c$ and is presented as a function of the normalized star cluster radius $R_{SC}/R_c$ (left panel), or normalized star cluster half-mass radius  $R_{hm}/R_{c}$ (right panel). In the case of top-hat profiles, with $\beta = 0$, we normalized to $R_c = 1$~pc. This makes the top-hat cluster radius or top-hat cluster half-mass radius dimensionless and comparable to other profiles with different values of $\beta$. There is a good correspondence between the results of semi-analytical and 1D hydrodynamical simulations. 

At low values of $R_{SC}/R_c$, where $R_{SC} <
R_{c}$, all the curves follow the critical luminosity of the top-hat profile
($\beta = 0$). At somewhat larger values of $R_{SC}/R_c$, critical luminosities for
models with $\beta > 0$ deviate from the top-hat line towards higher
luminosities. This is because their steeper stellar densities lead to the
fastly growing wind velocities and there is less time for cooling and clump formation. Therefore,
the cluster wind density and hence the luminosity must be larger to reach the
thermally unstable regime. At even higher values, $R_{SC}/R_c > 10$, the critical lines of models with $\beta \ge 1.5$ deviate
from the top-hat line towards smaller luminosities reaching a constant value
as $R_{SC} \rightarrow\infty$. This is because the denser central regions of these clusters undergoing rapid radiative cooling are not influenced by a further increase of $R_{SC}$. Critical luminosity curves for models with $0 <
\beta < 1.5$ deviate at large $R_{SC}/R_c$ from the top-hat critical line, however since the mass contribution of their peripheral parts is never negligible, they have even for $R_{SC}/R_c \rightarrow\infty$ some non-zero slope of the $L_{crit}/R_c$ vs $R_{SC}/R_c$ line depending on the value of $\beta$.

The $L_{crit}/R_c$ vs $R_{hm}/R_c$ profiles are shown in the right panel of Figure 6. They are similar to $L_{crit}/R_c$ vs $R_{SC}/R_c$ profiles.  
The difference appears at the high $R_{hm}/R_c$ values: for $\beta \ge 1.5$ the $R_{hm}/R_c$ can not grow to infinity. It has a finite value depending on $\beta$. For $\beta$ = 2 it is about 2.26 and for $\beta$ = 1.5 it is about 36.84.       

\section{Summary}

We have developed a model for the winds driven by stellar clusters with a
generalized Schuster stellar density distribution. Two methods: a
semi-analytic solution and 1D hydrodynamical simulations have been thoroughly
discussed and shown in excellent agreement for clusters whose mechanical
luminosities do not exceed the critical value, $L_{crit}$.

The semi-analytic solution cannot be applied for clusters with $L_{SC} >
L_{crit}$ as in this case radiative cooling is extremely fast and the solution
becomes thermally unstable.
 Nevertheless, assuming spherically symmetric star clusters, we have inferred the properties of stationary star cluster winds.
For example, we have shown that  in the adiabatic case, when radiative cooling is excluded, the position of the sonic point $R_{son}$, where the wind speed reaches the local sound speed, is given by the steepness of the stellar density distribution, if the singular point $R_{sg}$ of the equation (\ref{eq7a}) for the radial gradient of the wind speed is inside of the cluster: $R_{sg} = R_{son} < R_{SC}$. 
$R_{sg}$ is larger in clusters with flatter stellar  
density
distributions (smaller $\beta$) and   
goes to
infinity for $\beta \le \beta_{crit}$, where $\beta_{crit} = 9/8$.
This
implies that inside clusters with flat stellar density distributions the
flow is always subsonic and the transition to the supersonic flow  
occurs
at the star cluster surface, whereas in clusters with steeper density
distributions the transition to the supersonic regime may occur either
inside of the cluster, or at its surface, depending on whether the  
value of
the cut-off radius $R_{SC}$ is larger or smaller than that of the  
singular
point $R_{sg}$.

The position of the sonic point either inside the cluster or on its surface
leads to two kinds of very different winds: in the first case the wind velocity increases gradually from the cluster center, while in the shallow cases  there is a very slow subsonic wind inside  the cluster and a strong wind acceleration to supersonic speed just at the cluster surface. 

As one considers more massive clusters, the wind density increases implying a growing influence of radiative cooling: a fraction of the wind mechanical energy is lost and the wind temperature, velocity and sound speed decrease. The most energetic stationary winds exist for clusters with  moderate steepnesses ($\beta = 1 - 1.5$) although a large fraction of this energy is lost by radiation. For less or more steep stellar density profiles the wind becomes non-stationary at lower mechanical luminosities. The steepness of the stellar density profile also regulates the fraction of energy that is radiated away. Steeper stellar distributions lead to smaller sizes of the high density strongly radiative central regions with high wind speeds: there is less time for cooling of winds in clusters with steep stellar density profiles.

The dependence on the star cluster parameters, was explored.
The normalized critical energy, $L_{crit}/R_c$, was calculated as a function
of the normalized star cluster radius, 
$R_{SC}/R_c$
,  and also as a function
of the normalized star cluster half-mass radius, 
$R_{hm}/R_c$, 
and is
plotted in Figure 6. One can compare target cluster parameters
with these critical lines in order to find if radiative cooling may affect
the star cluster driven flow significantly.  
The position of the normalized critical line $L_{crit}/R_c$  
separates stationary winds from non-stationary winds in which
frequent thermal instabilities in the deposited matter lead to the rapid condensation of unstable parcels of gas  forming cold cloudlets  immersed in the pervasive  hot matter.

Clusters with a decreasing stellar density ($\beta > 0$) and $R_{SC} < 10 \times  R_c$ have $L_{crit}/R_c$ at higher values compared to the top-hat ($\beta = 0$) stellar density profiles. 
If $R_{SC} > 10 \times R_c$, in clusters with a steep stellar density distribution ($\beta \ge 1.5$), the critical luminosity $L_{crit}/R_c$ approaches a constant value because the central cooling regions are only marginally influenced by increasing $R_{SC}/R_c$.

The non-stationary winds formed in high mass clusters with the total luminosity above the critical value $L_{crit}$ will be explored using 3D hydrodynamical simulations in a forthcoming communication. 
We shall also investigate if there are clusters formed with a mass near the critical luminosity $L_{crit}$. The very compact MW cluster Arches is a candidate. Also some compact clusters in the LMC or in M82 and the Antennae galaxies may be close or above the critical line  suffering from strong internal cooling. 

\acknowledgments

The authors express their thanks to the anonymous referee, whose proposals have helped improve this paper.  
This study has been supported by the Czech Science Foundation grant 209/12/1795 and by the project RVO: 67985815; the Academy of Sciences of  the Czech Republic and  CONACYT-M\'exico research collaboration under the project 17048: Violent star formation;   the CONACYT - M\'exico, research grants 131913 and 167169.   
H-Z. F. wishes to express his thanks to CONACYT-M\'exico  for additional support  through grants 162184 and 186720. 

\bibliographystyle{aa}
\bibliography{Schuster-I}

\end{document}